\documentclass[aps,pre,10pt,showpacs,floatfix,onecolumn,nofootinbib,a4paper]{revtex4-2}
\usepackage{amssymb, amsmath, amsthm, mathtools}
\usepackage{bm}
\usepackage{mathrsfs}
\usepackage{hyperref,bookmark}
\usepackage{graphicx}
\usepackage{epsfig,color}
\usepackage{mathrsfs}
\usepackage{verbatim}
\usepackage{url}
\usepackage{subcaption}
\usepackage{float}
\usepackage[greek,english]{babel}
\usepackage{dcolumn}

\newcommand{\tr}{\operatorname{tr}}
\newcommand{\dd}{\operatorname{d}\!}

\newcommand{\diver}{\operatorname{div}}
\newcommand{\curl}{\operatorname{curl}}

\newcommand{\n}{\bm{n}}
\newcommand{\e}{\bm{e}}

\newcommand{\normal}{\bm{\nu}}
\newcommand{\body}{\mathscr{B}}
\newcommand{\free}{\mathscr{F}}
\newcommand{\boundary}{\partial\mathscr{B}}

\newcommand{\vt}{\vartheta}

\newcommand{\WOF}{W_\mathrm{OF}}
\newcommand{\WQT}{W_\mathrm{QT}}
\newcommand{\bend}{\bm{b}}

\newcommand{\Wn}{\mathbf{W}(\n)}
\newcommand{\Pn}{\mathbf{P}(\n)}
\newcommand{\Dn}{\mathbf{D}}
\newcommand{\I}{\mathbf{I}}
\newcommand{\twon}{(\n_1,\n_2)}
\newcommand{\zero}{\bm{0}}

\newcommand{\order}{\mathcal{O}}

\newcommand{\framec}{(\e_r,\e_\vt,\e_z)}
\newcommand{\degree}{^\circ}

\newcommand{\nigh}[1]{{\color{black}#1}}

\begin{document}
\latintext

\title{An Elastic  Quartic Twist Theory for Chromonic Liquid Crystals}
\author{Silvia Paparini}
\email{silvia.paparini@unipv.it}
\author{Epifanio G. Virga}
\email{eg.virga@unipv.it}
\affiliation{Dipartimento di Matematica, Universit\`a di Pavia, Via Ferrata 5, 27100 Pavia, Italy}

\begin{abstract}
Chromonic liquid crystals are lyotropic materials which are attracting growing interest for their adapatbility to living systems. To describe their elastic properties, the classical Oseen-Frank theory requires anomalously small twist constants and (comparatively) large saddle-splay constants, so large as to violate one of Ericksen's inequalities, which guarantee that the Oseen-Frank stored-energy density is bounded below. While such a violation does not prevent the existence and stability of equilibrium distortions in problems with fixed geometric confinement, the study of free-boundary problems for droplets has revealed a number of paradoxical consequences. Minimizing sequences driving the total energy to negative infinity have been constructed by employing ever growing needle-shaped tactoids incorporating a diverging twist [Phy. Rev. E \textbf{106}, 044703 (2022)]. To overcome these difficulties, we propose here a novel elastic theory that extends for chromonics the classical Oseen-Frank stored energy by adding a quartic twist term. We show that the total energy of droplets is bounded below in the quartic twist theory, so that the known paradoxes are ruled out. The quartic term introduces a phenomenological length $a$ in the theory; this affects the equilibrium of chromonics confined within  capillary tubes. Use of published experimental data allows us to estimate $a$.
\end{abstract}
\date{\today}

\maketitle

\section{Introduction}\label{sec:intro}
Liquid crystals are \emph{ordered} fluids divided into two big families: \emph{thermotropic} and \emph{lyotropic}. What distinguishes these families is the agent that drives the transition form ordinary, isotropic fluids to anisotropic ones, typically birefringent and optically uniaxial. It is temperature in the former family and concentration in the latter. Correspondingly, ordering takes place at sufficiently low temperature or sufficiently high concentration. 

Here we shall be concerned with the \emph{nematic} phase, where order arises only in the \emph{orientation} of the typical elongated constituents\footnote{These may be either molecules, molecular aggregates, or colloidal particles.} of liquid crystals, \emph{not} in their position in space. In both families, nematic order is described by the \emph{director} $\n$, a unit vector field which designates the local prevailing orientation, corresponding to the \emph{optic} axis. It can (easily) vary in space in response to different external agencies; its optical nature contributes to the aesthetic fascination of these materials.

When $\n$ has not everywhere the same orientation, its distorted \emph{texture} exhibits a \emph{curvature} elasticity described by a stored energy expressed in terms of the invariants of $\nabla\n$, the latter taken to be a tensorial measure of local distortion. The classical theory of nematic elasticity is based on the assumption that in the ground state $\n$ has everywhere the same orientation and that the energy stored in a distortion measures the work done to produce it starting from the ground state. The Oseen-Frank theory posits a stored energy quadratic in $\nabla\n$ and features four elastic constants, one for each elementary distortional mode.

Chromonic liquid crystals (CLCs) are lyotropic materials, which include Sunset Yellow (SSY), a popular dye in food industry, and disodium cromoglycate (DSCG), an anti-asthmatic drug. In these materials, molecules stuck themselves in columns, which in aqueous solutions develop a nematic orientational order. In CLCs, the director designates the average direction in space of the constituting  supra-molecular aggregates. 

A number of reviews have already appeared in the literature \cite{lydon:chromonic_1998,lydon:handbook,lydon:chromonic_2010,lydon:chromonic,dierking:novel}, to which we refer the reader. They also witness the scientific interest surrounding this theme.

\begin{figure}[h]
	\centering
	\begin{subfigure}[c]{0.4\linewidth}
		\centering
		\includegraphics[width=\linewidth]{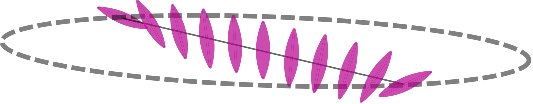}
		\caption{\nigh{Right}-handed} 
		\label{fig:ET_cartoon_left}
	\end{subfigure}
	\\
	\begin{subfigure}[c]{0.4\linewidth}
		\centering
		\includegraphics[width=\linewidth]{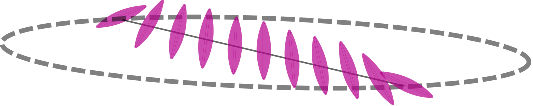}
		\caption{\nigh{Left}-handed} 
		\label{fig:ET_cartoon_right}
	\end{subfigure}
	\caption{Cartoons of two ET fields  with opposite chiralities in a cylinder. The nematic director $\n$ is schematized as a prolate ellipsoid. Conventionally, we call \nigh{right}-handed the arrangement in (a) as \nigh{the integral lines of $\n$ on the cylinder's lateral boundary are helices winding anti-clockwise around the cylinder's axis,} as the last four fingers of \nigh{your right hand would do around your right thumb oriented upward along the axis}. By contrast, \nigh{by the same convention}, the arrangement in (b) is then \nigh{left}-handed.}
	\label{fig:ET_cartoons}
\end{figure}

Experiments have been performed with these materials in capillary tubes, with either circular \cite{nayani:spontaneous,davidson:chiral} or rectangular \cite{fu:spontaneous} cross-sections, as well as on cylindrical shells \cite{javadi:cylindrical}, all enforcing \emph{degenerate planar} anchoring, which allows constituting columns to glide freely on the anchoring surface, provided they remain tangent to it. These experiments revealed a tendency of CLCs to acquire at equilibrium a \emph{twisted} configuration in cylinders, represented by an \emph{escaped twist} (ET) director field. Due to the lack of chirality in both the molecular aggregates constituting CLCs and their condensed phase, ET fields come equally likely in two variants with opposite chiralities, as illustrated by the cartoons in Fig.~\ref{fig:ET_cartoons}. \footnote{Escaped-twist (ET) is a name perhaps first used in in \cite{ondris-crawford:curvature} for what before had been called \emph{twist-bend}  in \cite{cladis:non-singular} or \emph{escape} in the third dimension in \cite{meyer:existence}.} 

Despite the lack of \emph{uniformity} in the ground state of these phases,\footnote{The  classification of the most general \emph{uniform} distortions, which can fill the whole three-dimensional space, is given in \cite{virga:uniform} and recalled in Sect.~\ref{sec:quadratic_formula}.} their curvature elasticity has been modeled by the  Oseen-Frank theory, albeit with an anomalously small twist constant $K_{22}$. To accommodate the experimental findings and justify the twisted ground state, this constant has to be smaller than the saddle-splay constant $K_{24}$, in violation of one of the inequalities Ericksen~\cite{ericksen:inequalities} had put forward to guarantee that the Oseen-Frank stored energy be bounded below.

Actually, as shown in \cite{paparini:stability}, the violation of one Ericksen's inequality does not prevent the twisted ground state from being locally stable in a cylinder enforcing degenerate planar anchoring on its lateral boundary. The same  conclusion was reached in \cite{long:violation}. But, as shown in \cite{paparini:paradoxes}, free-boundary problems may reveal noxious consequences of violating Ericksen's inequalities.   If $K_{22}<K_{24}$, a CLC droplet, tactoidal\footnote{\emph{Tactoids} are elongated, cylindrically symmetric shapes with pointed ends as poles.} in shape and surrounded by an isotropic fluid environment enforcing degenerate planar anchoring for the director at the interface, is predicted to be unstable against \emph{shape change}: it would split indefinitely in smaller tactoids while the total free energy plummets to negative infinity (see \cite{paparini:paradoxes}, for more details).

This prediction is in sharp contrast with the wealth of experimental observations of CLC tactoidal droplets, stable in the biphasic region of phase space, where nematic and isotropic phases coexist in equilibrium. Experiments have been carried out with a number of substances (including DSCG and SSY) stabilized by the addition of neutral (achiral) condensing agents (such as PEG and Spm) \cite{tortora:self-assembly,tortora:chiral,peng:chirality,nayani:using,shadpour:amplification}. These studies have consistently reported stable twisted bipolar tactoids.

Thus, we meet with a contradiction: if we adopt the Oseen-Frank theory to describe CLCs, we need to assume that $K_{24}>K_{22}$ to explain the twisted nematic textures observed  in cylindrical capillaries, but we also need to assume that $K_{24}<K_{22}$ to justify the very existence of twisted tactoids observed in the biphasic coexistence region. 

This paper presents our way to overcome this difficulty arising from applying the Oseen-Frank theory to the curvature elasticity of chromonics. We propose to add a specific \emph{quartic} term to the stored-energy density.

Higher-order theories are not new in liquid crystal science. Go under this name either theories that allow for higher spatial gradients of $\n$ in the energy and theories that allow for higher powers of the first gradient. Under the first category, which perhaps has seen its first manifestation in \cite{nehring:elastic} (see also \cite{oldano:ab}), falls, for example, Dozov's theory \cite{dozov:spontaneous} for both twist-bend and splay-bend phases predicted long ago by Meyer \cite{meyer:structural} and more recently observed in real materials \cite{cestari:phase}. Under the second category falls, for example, a simple one-dimensional model for splay-bend nematics \cite{lelidis:nematic}, then extended to incorporate a whole class of seven modulated ground states, of which twist-bend and splay bend are just two instances \cite{barbero:fourth}.
A hybrid theory was also proposed in \cite{lelidis:nonlinear}, where both higher gradients of $\n$ and higher powers of the first gradient are allowed in the stored-energy density, with spatial derivatives and their powers balanced according to a criterion motivated by a molecular model.\footnote{It should also be noted that other theories are known as ``quartic'' (see, for, example, the classical paper \cite{longa:extension} and the more recent contribution \cite{golovaty:novel}), but they owe this name to an elastic term globally quartic in deGennes' order tensor and its derivatives, added to the commonly considered version of the Landau-deGennes theory to resolve the spay-bend elastic constant degeneracy in the reduction to the Oseen-Frank theory. These theories serve a different purpose.}

The single quartic energy term contemplated in our theory will only contain the twist measure of nematic distortion; hence the name \emph{quartic twist} theory.

The paper is organized as follows. In Sect.~\ref{sec:energetics}, we describe the energetics of chromonics, starting from the classical quadratic Oseen-Frank theory. In particular,  we see what our theory has to say about the paradoxes for chromonic droplets encountered within the Oseen-Frank theory. Although we do not solve any free-boundary problem within the proposed theory, we provide a uniform lower bound for the total energy that prevents the existence of the minimizing sequences with diverging energy that generate paradoxes within the Oseen-Frank theory. In Sect.~\ref{sec:minimization_problem}, we apply the quartic twist theory to a CLC in a cylinder subject to degenerate planar anchoring on the lateral boundary and compare the predictions of this theory, for different values of a phenomenological length scale $a$ brought in by the added quartic term, with those of the classical theory, which is recovered in the limit where \nigh{$a$ vanishes.} 
Finally, in Sect.~\ref{sec:conclusion}, we summarize our conclusions. \nigh{In a closing Appendix, we collect the details of a fit of published data with our theory aiming at estimating both $K_{24}$ and $a$.}

\section{Energetics}\label{sec:energetics}
An extended theory may become more acceptable if its parent theory is first laid out. Thus, we start by summarizing the classical quadratic theory for the elasticity of nematic liquid crystals, albeit formulated in a novel, equivalent way that serves better our purpose. It will then become easier to justify how it should be modified to prevent the paroxysmal concentration of twist which in free-boundary problems leads the total energy to negative infinity \cite{paparini:paradoxes}. 

\subsection{Classical Quadratic Energy}\label{sec:quadratic_formula}
The classical elastic theory of liquid crystals goes back to the pioneering works of Oseen~\cite{oseen:theory} and Frank~\cite{frank:theory}.\footnote{Also a paper by Zocher~\cite{zocher:effect}, mainly concerned with the effect of a magnetic field on director distortions, is often mentioned among the founding contributions. Some authors go to the extent of also naming the theory after him. Others, in contrast, name the theory only after Frank, as they only deem his contribution to be fully aware of the nature of $\n$ as a \emph{mesoscopic}  descriptor of molecular order.} This theory is variational in nature, as it is based on a bulk free energy functional $\free_\mathrm{b}$ written in the form
\begin{equation}
	\label{eq:free_energy}
	\free_\mathrm{b}[\n]:=\int_{\body}\WOF(\n,\nabla\n)\dd V,
\end{equation}
where $\body$ is a region in space occupied by the material and $V$ is the volume measure. In \eqref{eq:free_energy}, $\WOF$ measures the distortional cost produced by a deviation from a uniform director field $\n$. It is chosen to be the most general frame-indifferent,  even function quadratic in $\nabla\n$, 
\begin{equation}
	\label{eq:free_energy_density}
	\begin{split}
		\WOF(\n,\nabla\n)&:=\frac{1}{2}K_{11}\left(\diver\n\right)^2+\frac{1}{2}K_{22}\left(\n\cdot\curl\n\right)^2+ \frac{1}{2}K_{33}\vert\n\times\curl\n\vert^{2} \\&+ K_{24}\left[\tr(\nabla\n)^{2}-(\diver\n)^{2}\right].
	\end{split}
\end{equation}
Here $K_{11}$, $K_{22}$, $K_{33}$, and $K_{24}$ are elastic constants characteristic of the material. They are traditionally referred to as the \emph{splay}, \emph{twist}, \emph{bend}, and \emph{saddle-splay} constants, respectively, by the features of four different orientation fields, each with a distortion energy proportional to a single term in \eqref{eq:free_energy_density} (see, for example, Chap.~3 of \cite{virga:variational}). 

Recently, Selinger~\cite{selinger:interpretation} has reinterpreted the classical formula \eqref{eq:free_energy_density} by decomposing the saddle-splay mode into a set of other independent modes. The starting point of this decomposition is a novel representation of $\nabla\n$ (see also \cite{machon:umbilic}),
\begin{equation}
	\label{eq:nabla_n_novel}
	\nabla\n=-\bend\otimes\n+\frac12T\Wn+\frac12S\Pn+\Dn,
\end{equation}
where $\bend:=-(\nabla\n)\n=\n\times\curl\n$ is the \emph{bend} vector, $T:=\n\cdot\curl\n$ is the \emph{twist}, $S:=\diver\n$ is the \emph{splay}, $\Wn$ is the skew-symmetric tensor that has $\n$ as axial vector, $\Pn:=\I-\n\otimes\n$ is the projection onto the plane orthogonal to $\n$, and $\Dn$ is a symmetric tensor such that $\Dn\n=\zero$ and $\tr\Dn=0$. By its own definition, $\Dn\neq\zero$ admits the following biaxial representation,
\begin{equation}
	\label{eq:D_representation}
	\Dn=q(\n_1\otimes\n_1-\n_2\otimes\n_2),
\end{equation}
where $q>0$ and $\twon$ is a pair of orthogonal unit vectors in the plane orthogonal to $\n$, oriented so that $\n=\n_1\times\n_2$.\footnote{It is argued in \cite{selinger:director} that $q$ should be given the name \emph{tetrahedral} splay, to which we would actually prefer \emph{octupolar} splay for the role played by a cubic (octupolar) potential on the unit sphere \cite{pedrini:liquid} in representing all scalar measures of distortion,  but $T$.}

By use of the following identity, 
\begin{equation}
	\label{eq:identity}
	2q^2=\tr(\nabla\n)^2+\frac12T^2-\frac12S^2,
\end{equation}
we can easily give \eqref{eq:free_energy_density} the equivalent form
\begin{equation}
	\label{eq:Frank_equivalent}
	\WOF(\n,\nabla\n)=\frac12(K_{11}-K_{24})S^2+\frac12(K_{22}-K_{24})T^2+\frac12K_{33}B^2+2K_{24}q^2,
\end{equation}
where $B^2:=\bend\cdot\bend$. Since $(S,T,B,q)$ are all independent \emph{distortion characteristics}, it readily follows from \eqref{eq:Frank_equivalent} that $\WOF$ is positive semi-definite whenever
\begin{subequations}\label{eq:Ericksen_inequalities}
	\begin{eqnarray}
		K_{11}&\geqq& K_{24}\geqq0,\label{eq:Ericksen_inequalities_1}\\
		K_{22}&\geqq& K_{24}\geqq0, \label{eq:Ericksen_inequalities_2}\\
		K_{33}&\geqq&0,\label{eq:Ericksen_inequalities_3}
	\end{eqnarray}
\end{subequations}
which are the celebrated \emph{Ericksen's inequalities} \cite{ericksen:inequalities}. If these inequalities are satisfied in strict form, the global ground state of $\WOF$ is attained on the uniform director field, characterized by
\begin{equation}
	\label{eq:uniform_ground_state}
	S=T=B=q=0.
\end{equation}
As already mentioned in the Introduction, inequality \eqref{eq:Ericksen_inequalities_2} must be violated for the ground state of $\WOF$ to be different from \eqref{eq:uniform_ground_state}, involving a non-vanishing $T$. 

The class of \emph{uniform} distortions was defined in \cite{virga:uniform} as the one comprising all director fields for which the distortion characteristics are constant in space. Equivalently said, a uniform distortion is a director field that can \emph{fill} three-dimensional space. It was proven that there are two distinct families of uniform distortions, characterized by the following conditions \cite{virga:uniform},
\begin{equation}\label{eq:uniform_distortions}
	S=0,\quad T=\pm2q,\quad b_1=\pm b_2=b,	
\end{equation} 
where $q$ and $b$ are arbitrary parameters such that  $q>0$ and $B^2=2b^2$, and $b_1$ and $b_2$ are defined by the decomposition
\begin{equation}
	\label{eq:b_1_b_2}
	\bend=b_1\n_1+b_2\n_2.
\end{equation}

The general director field corresponding to \eqref{eq:uniform_distortions} is the \emph{heliconical} ground state of twist-bend nematic phases,\footnote{With opposite \emph{chiralities}, one for each sign in \eqref{eq:uniform_distortions}.} in which $\n$ makes a fixed \emph{cone} angle with a given axis in space (called the \emph{helix} axis), around which $\n$ precesses periodically \cite{virga:uniform}.\footnote{In opposite senses, according to the sign of chirality.} The special instance in which $B=0$ corresponds to the \emph{single twist} that characterizes \emph{cholesteric} liquid crystals. 

The distortion for which all characteristics vanish, but $T$, is a \emph{double twist}.\footnote{Here we adopt the terminology proposed by Selinger~\cite{selinger:director} (see also \cite{long:explicit}) and distinguish between \emph{single} and \emph{double} twists, the former being \emph{uniform} and the latter not.} It is \emph{not} uniform and cannot fill space; it can possibly be realized locally, but not everywhere. In words, we say that it is a \emph{frustrated} ground state. It is, however, relevant to CLCs, as a double twist is attained exactly on the symmetry axis of both chiral variants of the ET field \cite{paparini:stability}. 

Liquid crystals are (within good approximation) incompressible fluids. Thus, when the region $\body$ is \emph{not} fixed, for a given amount of material, $\body$ is subject to
the  \emph{isoperimetric} constraint that prescribes its volume,
\begin{equation}
	\label{eq:isoperimentric_constraint}
	V(\body)=V_0.
\end{equation}
When $\body$ is surrounded by an isotropic fluid, a surface energy arises at the free interface $\boundary$, which, following \cite{rapini:distortion}, we represent as
\begin{equation}
	\label{eq:surface_energy}
	\free_\mathrm{s}[\n]:=\int_{\boundary}\gamma[1+\omega(\n\cdot\normal)^2]\dd A,
\end{equation}
where $\normal$ is the outer unit normal to $\boundary$, $\gamma>0$ is the \emph{isotropic} surface tension, $\omega>-1$ is a dimensionless parameter weighting the \emph{anisotropic} component of surface tension against the isotropic one, and $A$ denotes the area measure. For $\omega>0$, $\free_\mathrm{s}$ promotes the \emph{degenerate planar}  anchoring, whereas for $\omega<0$, it promotes the \emph{homeotropic} anchoring. For free-boundary problems, the total free energy functional will then be written as
\begin{equation}\label{eq:free_energy_total}
	\free_\mathrm{t}[\n]:=\free_\mathrm{b}[\n]+\free_\mathrm{s}[\n]
\end{equation}
and the domain $\body$, subject to \eqref{eq:isoperimentric_constraint}, is also an unknown to be determined so as to minimize $\free_{\mathrm{t}}$.

\subsection{Quartic Twist Energy}\label{sec:quartic_formula}
The essential feature of our theory is to envision a double twist with two equivalent chiral variants as ground state of CLCs in three-dimensional space,
\begin{equation}
	\label{eq:double_twist}
	S = 0, \quad T = \pm T_0, \quad B = 0, \quad q = 0.
\end{equation}
The degeneracy of the ground  double twist  in \eqref{eq:double_twist} arises from the achiral nature of the molecular aggregates that constitute these materials, which is reflected in the lack of chirality of their condensed phases.

The elastic stored energy must equally penalize both ground chiral variants. Our minimalistic  proposal to achieve this goal is to add  a \emph{quartic twist} term to the Oseen-Frank stored-energy density:
\begin{equation}
	\label{eq:quartic_free_energy_density}
	\begin{split}
		\WQT(\n,\nabla\n)&=\frac{1}{2}(K_{11}-K_{24})S^2+\frac{1}{2}(K_{22}-K_{24})T^2+ \frac{1}{2}K_{23}B^{2}\\&+\frac{1}{2}K_{24}(2q)^2+ \frac14K_{22}a^2T^4,
	\end{split}
\end{equation}
where $a$ is a \emph{characteristic length}. Unlike $\WOF$ when \eqref{eq:Ericksen_inequalities_2} is violated, $\WQT$ is bounded below whenever 
\begin{subequations}\label{eq:new_inequalities}
	\begin{eqnarray}
		K_{11}&\geqq&K_{24}\geqq0,\label{eq:new_inequalities_1}\\
		K_{24}&\geqq&K_{22}\geqq0, \label{eq:new_inequalities_2}\\
		K_{33}&\geqq&0.\label{eq:new_inequalities_3}
	\end{eqnarray}
\end{subequations}
If these inequalities  hold, as we shall assume here, then $\WQT$ is minimum at the  degenerate double-twist \eqref{eq:double_twist}
characterized by
\begin{equation}
	\label{eq:T_0min}
	T_0:=\frac{1}{a}\sqrt{\frac{K_{24}-K_{22}}{K_{22}}}.
\end{equation}
As a consequence, $\WQT$ obeys the following uniform bound,
\begin{equation}
	\label{eq:bulk_energy_ bound}
	\WQT\geqq -\frac{1}{4a^2}\frac{(K_{24}-K_{22})^2}{K_{22}}.
\end{equation}	
\nigh{The parameter $a$ encodes the \emph{bare} length scale associated with the (double) twist $T_0$ favored locally by the ground state}.\footnote{In the elastic model proposed in \cite{virga:uniform} for twist-bend nematics, a quartic free energy was posited that admits as ground state either of two families of uniform heliconical fields with opposite chirality. There too, a length scale appears in the equilibrium pitch. The distortion state characterized by this  length is the same everywhere.} As to the physical size of such a length scale, it may fall in a wide range. While at the lower end we may place the persistence length of the molecular order, which characterizes the flexibility of CLC aggregates,\footnote{The persistence length of self-assembled flexible aggregates is a length over which unit vectors tangential to the aggregates lose correlation. For CLCs, it is estimated on the order of tens to hundreds of $\mathrm{nm}$ \cite{zhou:lyotropic}} the upper end is hard to make definite. We expect that $a$ would be exposed to the same difficulty that affects many (if not all) supramolecular structures in lyotropic systems. The most telling example is perhaps given by cholesteric liquid crystals, which give rise to a chiral structure (characterized by a single twist $T=\pm2q$) starting from chiral molecules. If the macroscopic pitch (measured by $\vert1/T\vert$) were determined by the molecular chirality,\footnote{\emph{Via} the naive geometric argument that represents chiral molecules as cylindrical screws and derives the pitch of their assemblies by close packing them so as to fit grooves with grooves.} it would result several orders of magnitude smaller than the observed ones.\footnote{For lyotropic cholesterics, the mismatch between microscopic and macroscopic pitches, which has recently received new experimental evidence in  systems of basic living constituents \cite{stanley:dna,tortora:chiral_2011}, is still debated. Interesting theories based on either molecular shape fluctuations \cite{harris:microscopic,harris:molecular} or surface charge patterns \cite{kornyshev:chiral} have met with some experimental disagreement \cite{grelet:what}.} Here, we shall treat $a$ as a phenomenological parameter, to be determined experimentally.

From now on, in the definition \eqref{eq:free_energy} for $\free_{\mathrm{b}}$ we shall replace $\WOF$ with $\WQT$. 

\subsection{Energy Lower Bound}\label{sec:drop}
Our quartic theory is built with the intent of curing the paradoxes encountered within the Oseen-Frank theory when handling free-boundary problems for chromonics.
As shown in \cite{paparini:paradoxes}, these paradoxes ultimately rely on the explicit construction of minimizing sequences of director fields and admissible shapes $\{\n_h,\body_h\}_{h\in\mathbb{N}}$ for the total free energy $\free_{\mathrm{t}}$ in \eqref{eq:free_energy_total}, over which $\free_{\mathrm{t}}$ diverges to negative infinity. Thus, in short, $\free_{\mathrm{t}}$ is proved to be \emph{unbounded} below in a class of free-boundary problems, if $\WOF$ is chosen to be the stored-energy density with \eqref{eq:Ericksen_inequalities_2} violated.\footnote{For the classical quadratic energy in a \emph{fixed} domain with degenerate planar anchoring conditions, boundedness and local stability of the bulk free energy $\free_{\mathrm{b}}$ were established in \cite{paparini:stability}, even when \eqref{eq:Ericksen_inequalities_2} is violated.}

Besides the violation of \eqref{eq:Ericksen_inequalities_2}, all minimizing sequences provided in \cite{paparini:paradoxes} have a common feature: they induce a high concentration of twist in needle-shaped droplets made sharper and sharper. Here we show that such a disruptive mechanism cannot be at work within the quartic twist theory. We are not in a position to prove that free-boundary problems are well-posed within this theory. Indeed, these are formidable problems already within the classical quadratic theory (see, for example, the early works \cite{virga:drops,lin:nematic} and the more recent contributions \cite{geng:two-dimensional,lin:isotropic-nematic}). We rather prove that $\free_{\mathrm{t}}$ is uniformly bounded below.

It is an easy consequence of \eqref{eq:bulk_energy_ bound} and \eqref{eq:isoperimentric_constraint} that
\begin{equation}
	\label{eq:bound_F_b}
	\free_{\mathrm{b}}[\n]\geqq-\frac{(K_{24}-K_{22})^2}{K_{22}}\frac{V_0}{4a^2}.
\end{equation}	
Similarly, it follows from \eqref{eq:surface_energy} that
\begin{equation}
	\label{eq:bound_F_s}
	\free_\mathrm{s}[\n]\geqq\gamma_\omega A(\partial\body),
\end{equation}
where
\begin{equation}
	\label{eq:gamma_omega}
	\gamma_\omega:=\gamma(1+\min\{0,\omega\})>0
\end{equation}
and $A(\boundary)$ is the area of $\boundary$. By the isoperimetric inequality (see, for example, \cite{osserman:isoperimetric}),
\begin{equation}
	\label{eq:isoperimetric_inequality}
	A(\partial\body)\geqq(36\pi V_0^2)^{1/3},
\end{equation}
and so \eqref{eq:bound_F_b} and \eqref{eq:bound_F_s} imply that 
\begin{equation}
	\label{eq:bound_F_t}
	\free_{\mathrm{t}}[\n]\geqq -\frac{(K_{24}-K_{22})^2}{K_{22}}\frac{V_0}{4a^2}+\gamma_\omega(36\pi V_0^2)^{1/3},
\end{equation}
thus establishing a lower bound for $\free_{\mathrm{t}}$ that only depends on material constants and the prescribed volume $V_0$.

\section{Chromonics in a Cylinder}\label{sec:minimization_problem}
Having established that the paradoxical behaviour encountered for chromonic droplets within the classical quadratic theory is ruled out by our quartic theory, we can study within the latter the problem that in \cite{nayani:spontaneous,davidson:chiral} first led to interpret as ET fields the equilibrium distortions observed in chromonics filling a cylinder enforcing degenerate planar anchoring conditions on its later boundary.

\begin{figure}[h]
	\centering 
	\includegraphics[width=.33\linewidth]{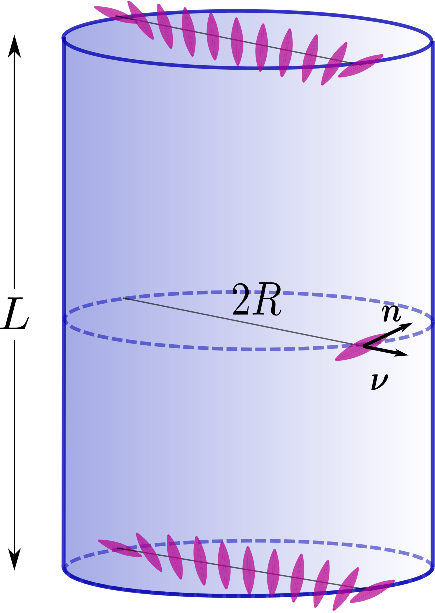}
	\caption{A circular cylinder of radius $R$ and height $L$ filled with a chromonic liquid crystal. The lateral boundary  $\partial_R\body$ enforces a degenerate planar anchoring condition for the director, here represented by a prolate ellipsoid with long axis orthogonal to the outer unit normal $\normal$, but free to rotate about it.}
	\label{fig:cylinder}
\end{figure}

Here the domain $\body$ is a circular cylinder of radius $R$ and height $L$ (see Fig.~\ref{fig:cylinder}) 
and the director field is subject to a degenerate planar anchoring condition on the lateral boundary  $\partial_R\body$ of $\body$:
\begin{equation}
	\label{eq:degenerate_planar_condition}
	\n\cdot\normal\vert_{\partial_R\body}=0,
\end{equation}
where $\normal$ denotes the outer unit normal to $\partial_R\body$. Free-boundary conditions are imposed on the bases of $\body$; the solution is assumed to be invariant under translations along the cylinder's axis.

As a consequence of \eqref{eq:degenerate_planar_condition}, $\free_\mathrm{s}$ in \eqref{eq:surface_energy} is a constant independent of $\n$ and will be omitted. The total free energy $\free_{\mathrm{t}}$ in \eqref{eq:free_energy_total} thus reduces to $\free_{\mathrm{b}}$ and will be scaled to the reference energy $2\pi K_{22}L$,
\begin{equation}
	\label{eq:free_energy_scaled}
	\mathcal{F}_\mathrm{b}[\n]:=\frac{\free_{\mathrm{b}}[\n]}{2\pi K_{22}L}.
\end{equation}

We further assume that in the frame $\framec$ of cylindrical coordinates $(r,\vt,z)$, with $\e_z$ along the axis of $\body$, $\n$ is represented as
\begin{equation}
	\label{eq:n_bur}
	\n=\sin\beta(r)\e_\vt+\cos\beta(r)\e_z,
\end{equation}
where the \emph{polar} angle $\beta\in[-\pi,\pi]$, which $\n$ makes with the cylinder's axis, depends only on the radial coordinate $r$. In these coordinates, $\partial_R\body$ is represented as the set $\{(r,\vt,z):r=R,0\leqq\vt\leqq2\pi,0\leqq z\leqq L\}$, and \eqref{eq:degenerate_planar_condition} is automatically satisfied, as $\normal\vert_{\partial_R\body}=\e_r$.
Moreover, as shown in \cite{paparini:stability}, the distortion characteristics for a director field represented by \eqref{eq:n_bur} are given by
\begin{subequations}\label{n_bur_characteristics}
	\begin{eqnarray}
		S&=&0,\label{n_bur_characteristics_S}\\
		T&=&\beta'+\frac{1}{r}\cos\beta\sin\beta,\label{n_bur_characteristics_T}\\
		B&=&\frac{1}{r}\sin^2\beta,\label{n_bur_characteristics_B}\\
		q&=&\frac12\left\lvert\beta'-\frac{1}{r}\cos\beta\sin\beta\right\rvert,\label{n_bur_characteristics_q}
	\end{eqnarray}
\end{subequations}
where a prime denotes differentiation. 

Standard computations transform $\mathcal{F}_\mathrm{b}$ in \eqref{eq:free_energy_scaled} into the \emph{reduced} functional
\begin{equation}
	\label{eq:reduced_free_energy}
	\mathcal{F}_\lambda[\beta]:=\mathcal{F}_2[\beta]+\lambda^2\mathcal{F}_4[\beta],
\end{equation}
where $\mathcal{F}_2$ corresponds to the (scaled) quadratic  Oseen-Frank's free energy functional, 
\begin{equation}
	\label{eq:free_T2}
	\begin{split}
		\mathcal{F}_2[\beta]&:= \int_0^1\left(\frac{\rho\beta'^2}{2}+\dfrac{1}{2\rho}\cos^2\beta\sin^2\beta+\dfrac{k_3}{2\rho}\sin^4\beta\right) \dd\rho\\&+\frac12(1-2k_{24})\sin^2\beta(1),
	\end{split}
\end{equation}
while
\begin{equation}
	\label{eq:free_k4}
	\mathcal{F}_{4}[\beta]:=\int_0^1\frac{\rho}{4}\left(\beta'+\frac{1}{\rho}\sin\beta\cos\beta\right)^4 \dd\rho,
\end{equation}
is the quartic contribution. Here
\nigh{
	\begin{equation}
		\label{eq:lambda_definition}
		\lambda:=\frac{a}{R},
\end{equation}} 
\begin{equation}
	\label{eq:rho_definition}
	\rho:=\frac{r}{R}\in[0,1],
\end{equation}
and 
\begin{equation}
	\label{eq:scaled_elastic_constant}
	k_3 :=\frac{K_{33}}{K_{22}}, \quad k_{24}:=\frac{K_{24}}{K_{22}}
\end{equation}
are scaled elastic constants, which will be assumed to obey the inequalities
\begin{equation}
	\label{eq:scaled_constants_inequalities}
	k_3>0\quad\text{and}\quad k_{24}>1,
\end{equation}
the strong form of \eqref{eq:new_inequalities_3} and \eqref{eq:new_inequalities_2}, respectively.

For the integral in \eqref{eq:reduced_free_energy} to be convergent, $\beta$ must be subject to the condition 
\begin{equation}
	\label{eq:boundary_condition_0}
	\beta(0)=0,
\end{equation}
which amounts to require that $\n$ is along $\e_z$ on the cylinder's axis.\footnote{\nigh{Alternatively, we could require $\beta(0)=\pm\pi$, but this would amount to change $\n$ into $-\n$, which does not affect the energy.}} The strong form of the Euler Lagrange equation for $\mathcal{F}_\lambda$ in \eqref{eq:reduced_free_energy} reduces here to
\begin{align}
	\label{eq:EL_bulk_clcs}
	\frac{1}{\rho^2}\cos\beta\sin\beta&\left[1+2(k_3-1)\sin^2\beta\right]-\frac{\beta'}{\rho}-\beta''\nonumber\\
	+ \lambda^2\left(\beta'+\frac{\sin\beta\cos\beta}{\rho}\right)^2&\left[ \frac{4\sin^2\beta\beta'}{\rho}-\frac{3\beta'}{\rho}\right.\nonumber\\
	&-\left.3\beta''+\frac{\sin\beta\cos\beta}{\rho^2}\left(3-2\sin^2\beta\right)\right]=0,
\end{align}
subject to \eqref{eq:boundary_condition_0} and to the following free-boundary condition at $\rho=1$,
\begin{equation}
	\label{eq:boundary_condition_1}
	\left[(1-2k_{24})\cos\beta\sin\beta+\beta'+\lambda^2\left(\beta'+\sin\beta\cos\beta\right)^3\right]_{\rho=1}=0.
\end{equation}
Solution to \eqref{eq:EL_bulk_clcs} subject to \eqref{eq:boundary_condition_0} and \eqref{eq:boundary_condition_1} enjoy a \emph{chiral} symmetry: if $\beta(\rho)$ is a solution, then also $-\beta(\rho)$ is so, which by \eqref{n_bur_characteristics} shares with the former all distortion characteristics, but $T$, which is opposite in sign. All chirality-conjugated solutions have the same energy, as follows immediately from \eqref{eq:reduced_free_energy} with \eqref{eq:free_T2} and \eqref{eq:free_k4}. For definiteness, in the following, we shall resolve this degeneracy by taking $\beta\geqq0$; it should always be taken as also representing its conjugated companion with opposite twist.\footnote{We are tacitly assuming that $\beta$ has only one sign in $[0,1]$, which is a property expected for the minimizers of $\mathcal{F}_\lambda$.} 

For $\lambda=0$, equations \eqref{eq:EL_bulk_clcs}, \eqref{eq:boundary_condition_0}, and \eqref{eq:boundary_condition_1} have a closed-form solution, given by (see \cite{burylov:equilibrium} and also \cite{davidson:chiral,paparini:stability})
\begin{equation}
	\label{eq:bur_solution}
	\beta_0(\rho):=\arctan\left(\frac{2\sqrt{k_{24}(k_{24}-1)}\rho}{\sqrt{k_{3}}\left[k_{24}-(k_{24}-1)\rho^2\right]}\right).
\end{equation}
Unfortunately, for $\lambda>0$, the solution is not known explicitly. We now discuss some numerical solutions for various values of $\lambda$ and then study analytically their asymptotics. 

\subsection{Numerical Solutions}\label{sec:numercial_solutions}
To illustrate our theory, we solve numerically equation \eqref{eq:EL_bulk_clcs} subject to \eqref{eq:boundary_condition_0} and \eqref{eq:boundary_condition_1} using the experimental data for DSCG available from \cite{zhou:elasticity_2012} and \cite{eun:effects}. The latter refer to an aqueous solution with concentration $c=14.0\%$ (wt/wt) at temperature $21.5\,\degree\mathrm{C}$, for which $k_3=30$ and $k_{24}=7.5$.

These data were obtained by using the Oseen-Frank theory. We argue that the experimental determination in \cite{zhou:elasticity_2012} of $K_{11}$, $K_{22}$, and $K_{33}$ (and so of $k_3$) would \emph{not} be affected by the quartic twist term introduced here in the elastic stored-energy density because those measurements result from an instability driven by a magnetic field; the occurrence of such an instability is determined by the change in sign of the second variation of the total free-energy functional. The quartic twist term does \emph{not} affect the second variation, and thus neither does the quartic elastic constant the magnetic instability. On the other hand, in our theory, measurements of $K_{24}$ are expected to reveal the role played by the quartic twist term.

For DSCG, the data available from \cite{eun:effects} do not allow us to estimate both $k_{24}$ and $\lambda$, and so to illustrate the role of $\lambda$, we first adopt the value of $k_{24}$ derived from the quadratic theory. Below, in Sect.~\ref{sec:estimates}, we use more abundant data available for SSY to estimate both $k_{24}$ and $\lambda$ for that material using our quartic theory.

\begin{figure}[h]
	\centering
	\includegraphics[width=.5\linewidth]{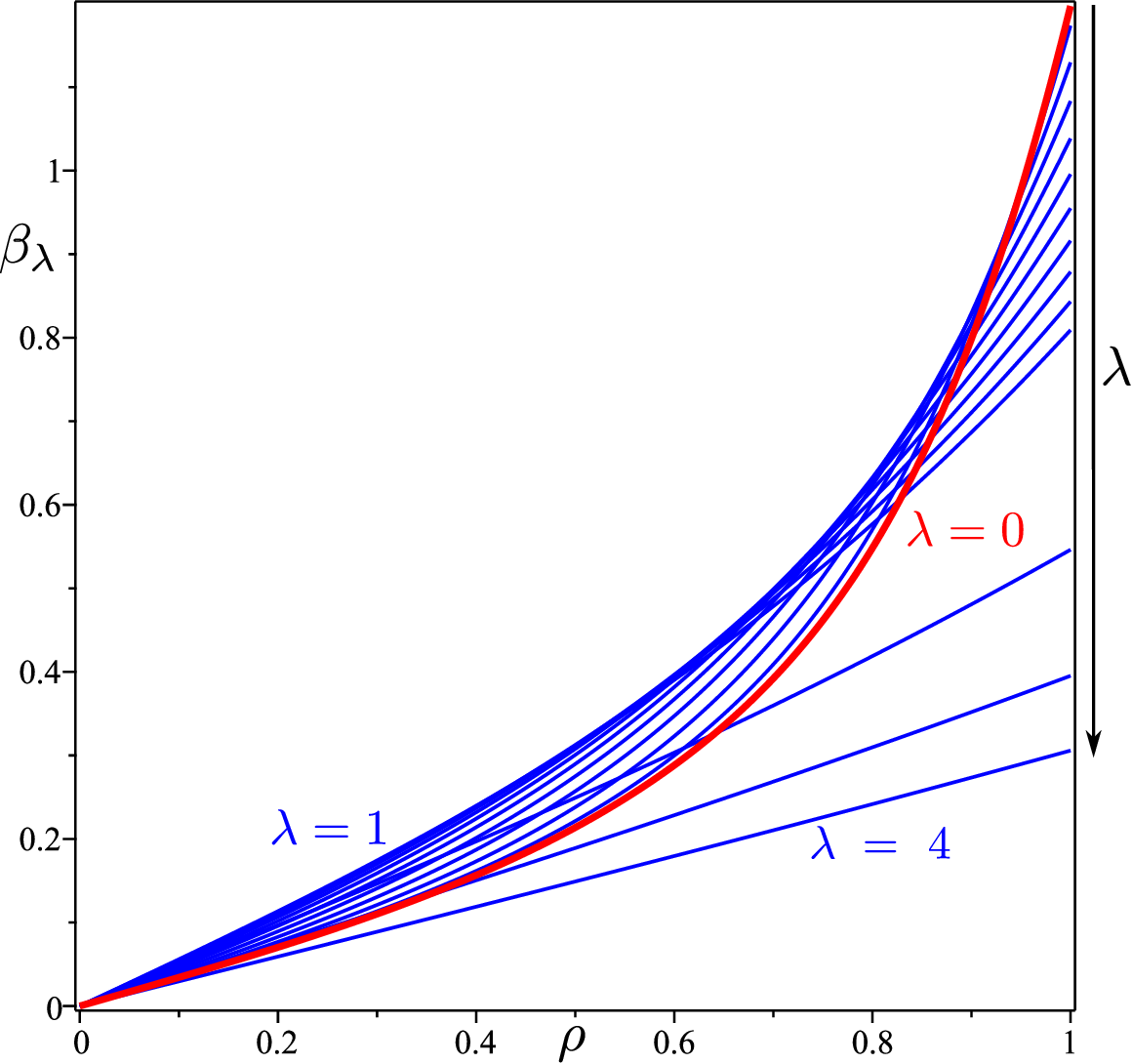}
	\caption{Graphs of the solution $\beta_\lambda(\rho)$ of \eqref{eq:EL_bulk_clcs} subject to \eqref{eq:boundary_condition_0} and \eqref{eq:boundary_condition_1} for several values of $\lambda$: ten  equally spaced between $\lambda=0$ and $\lambda=1$, followed by $\lambda=2,3,4$. The (red) graph corresponding to $\lambda=0$ reproduces the solution $\beta_0$ in \eqref{eq:bur_solution}. Material moduli are such that $k_3=30$ and $k_{24}=7.5$; they are taken from \cite{zhou:elasticity_2012} and \cite{eun:effects} and refer to an aqueous solution of DSCG with concentration $c=14\%$ (wt/wt) at $21.5\,\degree\mathrm{C}$.}
	\label{fig:beta_profiles}
\end{figure}

Figure~\ref{fig:beta_profiles} shows the graphs of the solution $\beta_\lambda(\rho)$ for $\lambda$ ranging from $0$ to $4$. For $\lambda=0$ (red graph in Fig.~\ref{fig:beta_profiles}), $\beta_\lambda$ reduces to $\beta_0$ in \eqref{eq:bur_solution}. As $\lambda$ increases, the convexity of the graph becomes less and less pronounced, and it is almost lost for $\lambda=4$, where the linear asymptotic regime described in Sect.~\ref{sec:asymptotics} below is already nearly attained. 

\begin{figure}[h]
	\centering
	\begin{subfigure}[c]{0.49\linewidth}
		\centering
		\includegraphics[width=\linewidth]{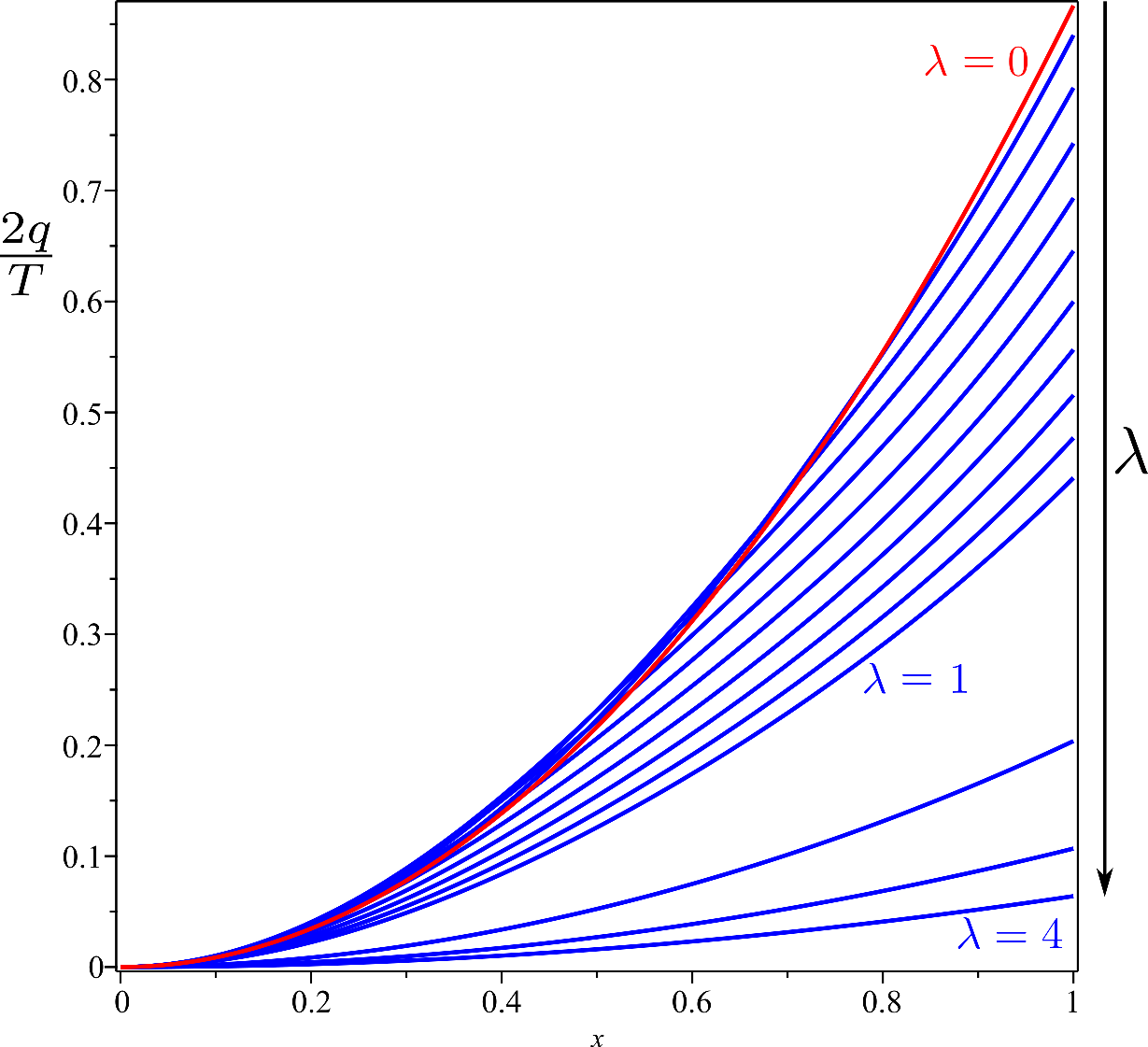}
		\caption{Uniformity ratio.}
		\label{fig:uniformity_ratio}
	\end{subfigure}
	%$\quad$
	\begin{subfigure}[c]{0.49\linewidth}
		\centering
		\includegraphics[width=\linewidth]{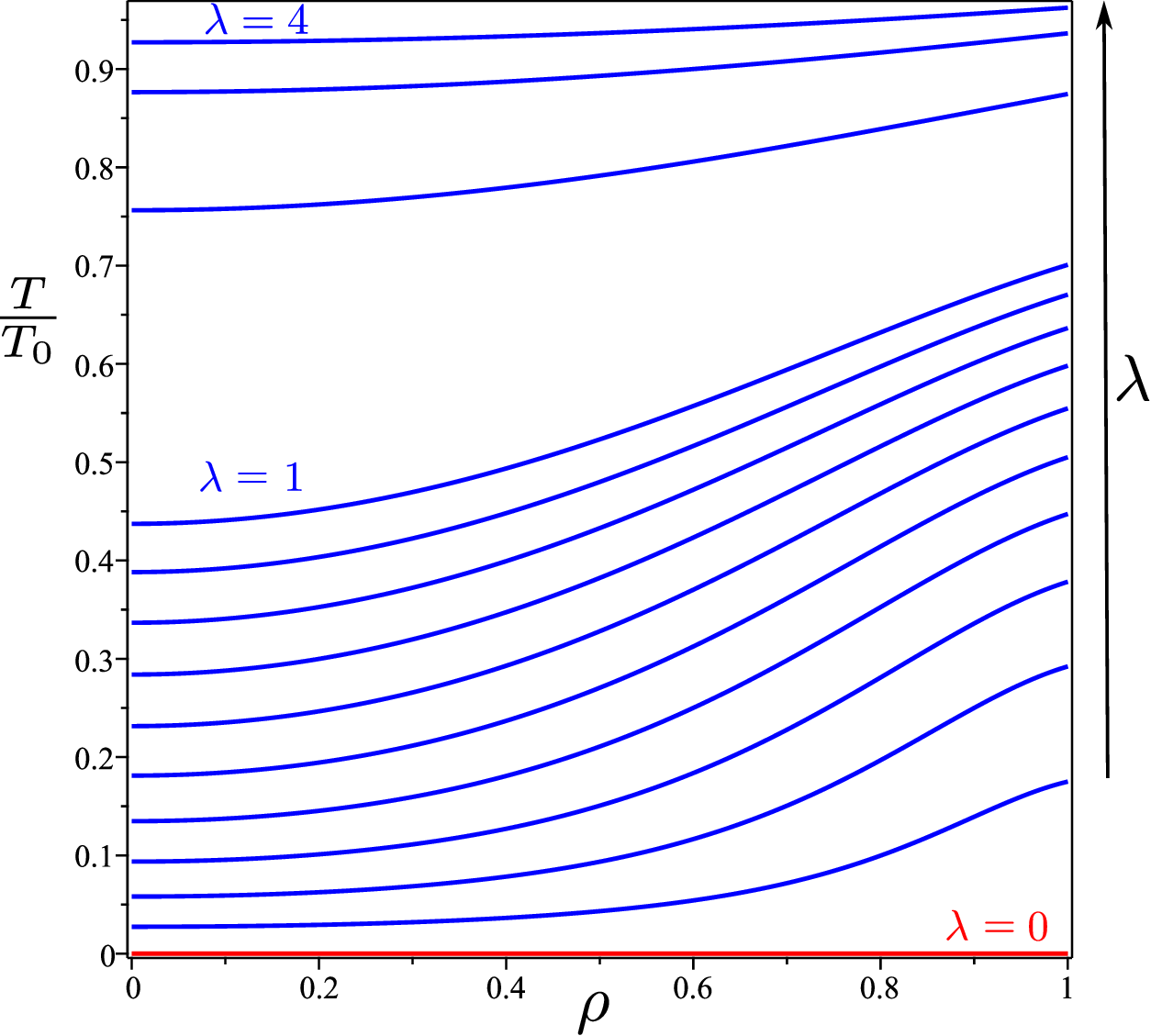}
		\caption{Frustration ratio.}
		\label{fig:frustration_ratio}
	\end{subfigure}
	\caption{Graphs of both uniformity and frustration ratios in \eqref{eq:uniformity_frustration_ratios} as functions of $\rho$. All parameters are the same as in Fig.~\ref{fig:beta_profiles}. Red curves refer to the analytic solution $\beta_0$ in \eqref{eq:bur_solution}, for which $\lambda=0$. The corresponding frustration ratio vanishes identically because $T_0$ diverges to $+\infty$ as $\lambda\to0$.}
	\label{fig:ratios}
\end{figure}

To further illustrate these solutions, we introduce the following dimensionless ratios,
\begin{subequations}
	\label{eq:uniformity_frustration_ratios}
	\begin{align}
		\frac{2q}{T}&=\frac{\beta_\lambda'(\rho)-\frac{1}{\rho}\cos\beta_\lambda(\rho)\sin\beta_\lambda(\rho)}{\beta_\lambda'(\rho)+\frac{1}{\rho}\cos\beta_\lambda(\rho)\sin\beta_\lambda(\rho)},\label{eq:uniformity_ratio}\\
		\frac{T}{T_0}&=\lambda\frac{\beta_\lambda'(\rho)+\frac{1}{\rho}\cos\beta(\rho)\sin\beta(\rho)}{\sqrt{k_{24}-1}},\label{eq:frustration_ratio}
	\end{align}	
\end{subequations}
where use was also made of both \eqref{n_bur_characteristics} and \eqref{eq:T_0min}. They are functions of $\rho$, which may be taken to represent local measures of \emph{uniformity} and \emph{frustration}, respectively. Where $2q/T$ gets closer to $1$, the distortion is nearer to be uniform, as intended in \eqref{eq:uniform_distortions}, and so fitter  to fill space. Where $T/T_0$ gets closer to $1$, the distortion is farther from being frustrated, as the energy is closer to its absolute minimum. The graphs of both these ratios are plotted in Fig.~\ref{fig:ratios} for different values of $\lambda$.
Upon increasing $\lambda$, the distortion in the cylinder becomes both less uniform and less frustrated.

\begin{figure}[h]
	\centering
	\includegraphics[width=0.5\linewidth]{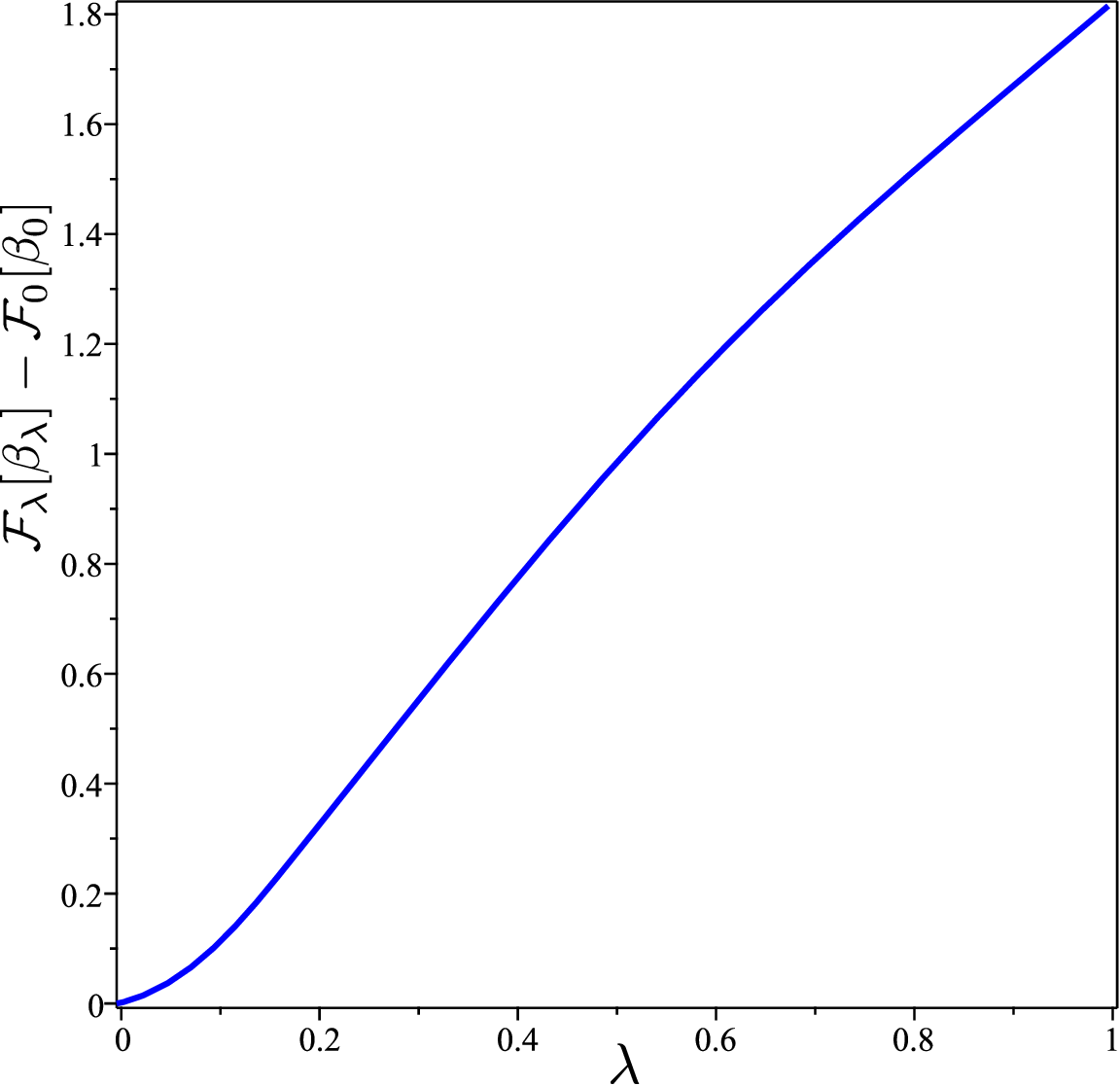}
	\caption{Graph of the minimum total energy $\mathcal{F}_\lambda[\beta_\lambda]$ against $\lambda$. Its behaviour for $\lambda\to0$ is in agreement with the asymptotic analysis in Sect.~\ref{sec:asymptotics}.}
	\label{fig:energy_profile}
\end{figure}

Finally, in Fig.~\ref{fig:energy_profile}, we plot the minimum of the total energy $\mathcal{F}_\lambda[\beta_\lambda]$ (referred to $\mathcal{F}_0[\beta_0]$) against $\lambda$. The asymptotic behaviour shown by this graph as $\lambda\to0$ will be justified in Sect.~\ref{sec:asymptotics} below. 

\subsubsection{Experimental Estimate}\label{sec:estimates}
\begin{figure}[h] 
	\centering
	\includegraphics[width=0.55\linewidth]{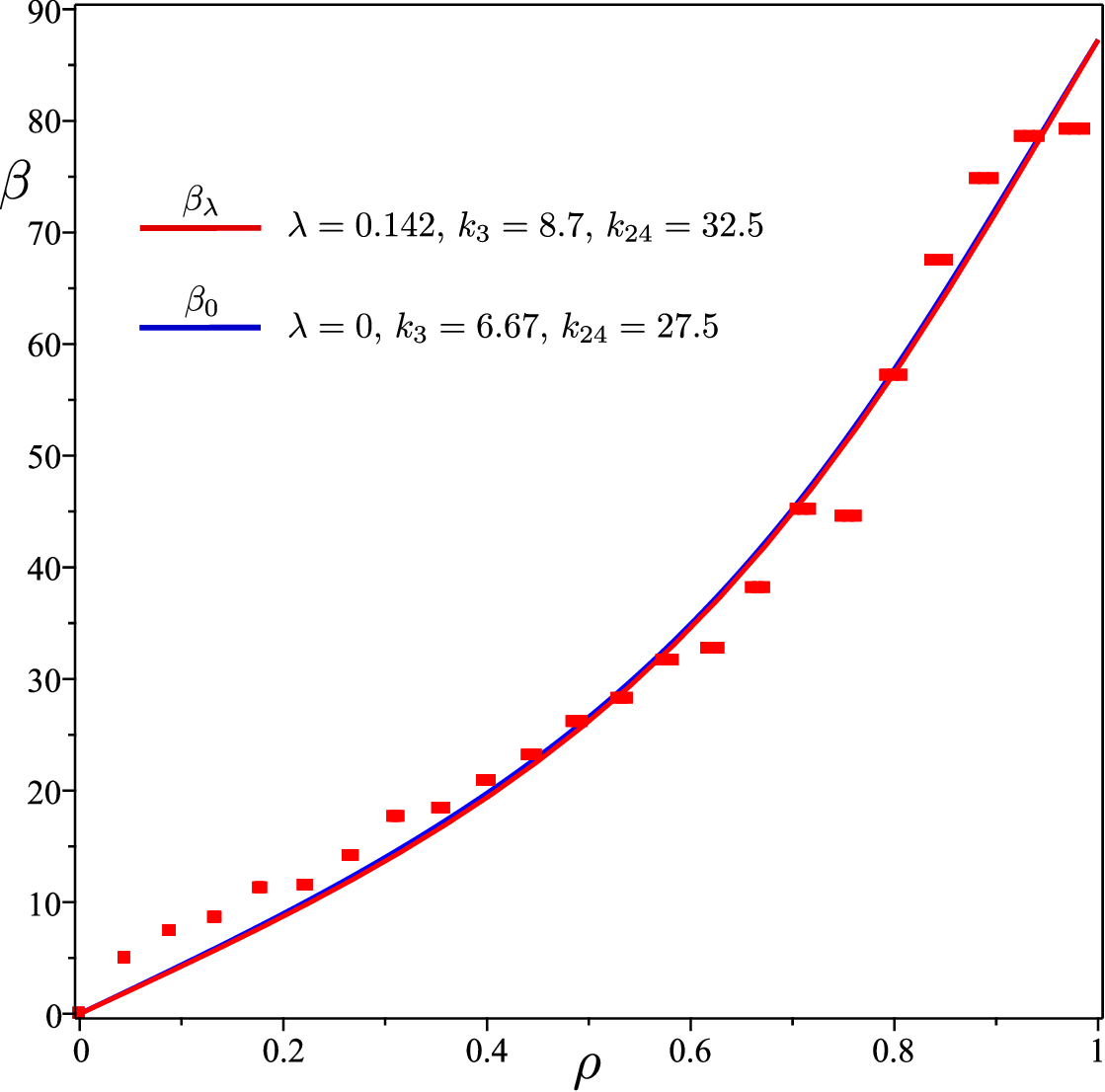}
	\caption{Best fit of the data shown in Fig.~3 of \cite{davidson:chiral} with the solution $\beta_\lambda$ to \eqref{eq:EL_bulk_clcs} for $k_{3}=8.7$ taken from \cite{zhou:elasticity_2012} \nigh{(red curve)}. Best fitting parameters were found to be  \nigh{$k_{24}=32.5$ and $\lambda=0.142$. The best fit of the same data with $\beta_0$ in \eqref{eq:bur_solution} for $\lambda=0$ and \emph{both} $k_3$ and $k_{24}$ fitting parameters is also shown (blue curve). The best fitting parameters were found to be $k_3=6.67$ and $k_{24}=27.5$, in agreement with \cite{davidson:chiral}. The two fitting curves are hardly discernible.}}
	\label{fig:estimate}.
\end{figure}

In \cite{davidson:chiral}, an experiment was performed with a solution of SSY in water (at a concentration $c=30.0\%$ (wt/wt) and temperature $25\,\degree\mathrm{C}$) inserted in a capillary tube with \nigh{radius $45.3\,\mu\mathrm{m}\pm400\,\mathrm{nm}$}. In equilibrium, the polar angle $\beta$ was measured at different distances from the capillary's axis; these data (taken from Fig.~3 of \cite{davidson:chiral}) are reproduced in Fig.~\ref{fig:estimate} \nigh{(red curve)}.
\nigh{We regard this particular experiment as  a case study where to put our theory to the test and show how both $K_{24}$ and $a$ could be determined. A systematic collection of data would be required to achieve a sound experimental determination of theses physical parameters.}

Taking $k_3=8.7$ from the measurements performed in \cite{zhou:elasticity_2012} for the same material in the same conditions,\footnote{The absolute measured values are $K_{11}\approx4.3\,\mathrm{pN}$, $K_{22}\approx0.7\,\mathrm{pN}$, and $K_{33}\approx6.1\,\mathrm{pN}$.} we determined $k_{24}$ and $\lambda$ from the best (least squares) fit shown in Fig.~\ref{fig:estimate}, obtaining
\begin{equation}
	\label{eq:fit_outcomes}
	\nigh{k_{24}=32.5\quad\text{and}\quad\lambda=0.142,}
\end{equation}
from which we estimate
\begin{equation}
	\label{eq:fit_estimates}
	\nigh{K_{24}\approx22.8\,\mathrm{pN}\quad\text{and}\quad a\approx6.4\,\mu\mathrm{m}.}
\end{equation}

\nigh{The details of our fit are given in Appendix~\ref{sec:fits}. Here we just heed that ours differs from the fit performed in \cite{davidson:chiral} (blue curve in Fig.~\ref{fig:estimate}) under an important respect. We have taken the bulk moduli $K_{22}$ and $K_{33}$ (and hence $k_3$) as determined in \cite{zhou:elasticity_2012} by customary bulk instabilities, whereas in \cite{davidson:chiral} $k_3$ was itself regarded as a fitting parameter in the analytic formula for the polar angle in \eqref{eq:bur_solution}. We consider the bulk determination of the bulk moduli more reliable.
	
	We have also explored the consequences of taking $k_3=8.7$ as in \cite{zhou:elasticity_2012} and $\lambda=0$: the best fit of the data in Fig.~\ref{fig:estimate} with $\beta_0$ in \eqref{eq:bur_solution} was then found for $k_{24}=222$, which is unrealistic. This negative result, however, shows how sensitive fitting these data is to the choice of $k_3$.}\footnote{\nigh{This was further witnessed by checking how the fitting minimum error obtained for the parameters in \eqref{eq:fit_outcomes} could be reduced by slightly reducing $k_3$ from the chosen value $k_3=8.7$.}}

\nigh{As shown in Fig.~\ref{fig:estimate}, for this particular experiment, the fits for the two theories compared here are hardly discernible. Both theories and fits, however, differ substantially in their premises: the theories differ in their energies, the fits differ in their choices for $k_3$.
	
	The papers \cite{davidson:chiral} and \cite{nayani:spontaneous} have provided estimates of the ratio $K_{24}/K_{33}$ for SSY. To compare these estimates to ours, which by \eqref{eq:fit_estimates} is $K_{24}/K_{33}\approx3.7$, we should remark that, due to a different scaling of the saddle-splay energy, the following conversion formulas apply
	\begin{equation}
		\label{eq:conversion}
		K_{24}=\frac12K'_{24}\quad\text{and}\quad K_{24}=\frac12(K''_{24}+K_{22}),
	\end{equation}	
	where $K'_{24}$ and $K''_{24}$ are the saddle-splay constants according to \cite{davidson:chiral} and \cite{nayani:spontaneous}, respectively. Thus, $K'_{24}/K_{33}$ in the range $(3.8,9.4)$ becomes $K_{24}/K_{33}$ in the range $(1.9,4.7)$ and $K''_{24}/K_{33}\approx6.7$ becomes $K_{24}/K_{33}\approx3.4$. Our estimate seems to be in good agreement with these.  
}
\subsection{Asymptotics}\label{sec:asymptotics}
Here, to justify the outcomes of the numerical computations shown above, we study the asymptotic behaviour of $\mathcal{F}_\lambda$ and its minimizers for both $\lambda\to0$ and $\lambda\to\infty$.

We first take $\lambda\ll1$ and assume that in this limit $\beta_\lambda$ is a perturbation to $\beta_0$, the minimizer of $\mathcal{F}_0$ described by \eqref{eq:bur_solution},
\begin{equation}\label{eq:beta_perturbation_0}
	\beta_\lambda(\rho)=\beta_0(\rho)+\lambda^\alpha u(\rho),
\end{equation}
where $\alpha>0$ is a scalar parameter and $u$ is a function in
in $\mathcal{H}^{1}([0,1])\cap \mathcal{L}^\infty([0,1])$ subject to $u(0)=0$, while remaining free at $\rho=1$. We shall determine \emph{both} $\alpha$ and $u$ by minimizing the approximate form of $\mathcal{F}_\lambda$ evaluated at $\beta_\lambda$ in \eqref{eq:beta_perturbation_0}.

To this end, we compute 
\begin{equation}
	\label{eq:F_lambda_0}
	\mathcal{F}_\lambda[\beta_\lambda]\approx\mathcal{F}_0[\beta_0]+\lambda^\alpha\delta\mathcal{F}_0(\beta_0)[u]+\lambda^2\mathcal{F}_4[\beta_0]+\frac12\lambda^{2\alpha}\delta^2\mathcal{F}_0(\beta_0)[u]+\lambda^{2+\alpha}\delta\mathcal{F}_4(\beta_0)[u],
\end{equation}
where $\delta\mathcal{F}_0(\beta_0)$ and $\delta^2\mathcal{F}_0(\beta_0)$ are the first and second variations of $\mathcal{F}_0$ at $\beta_0$, while $\delta\mathcal{F}_4(\beta_0)$ is the first variation of $\mathcal{F}_4$. Since $\beta_0$ is a critical point of $\mathcal{F}_0$, $\delta\mathcal{F}_0(\beta_0)\equiv0$. Since, as proved in \cite{paparini:stability}, $\delta^2\mathcal{F}_0(\beta_0)$ is positive definite, the only minimizer of $\mathcal{F}_\lambda$ in \eqref{eq:F_lambda_0} is trivially $u\equiv0$ as long as $\alpha<2$. For $\alpha=2$, $u$ is determined by minimizing the  quartic term in $\lambda$, the least order depending on $u$, which is a well-behaved functional with both quadratic and linear terms in $u$. The quadratic term in $\lambda$ already dominates  the energy $\mathcal{F}_\lambda$ and is independent of $u$. For $\alpha>2$, the last term in \eqref{eq:F_lambda_0} is dominant, but the minimizing $u$ does not exist, as the functional
\begin{equation}
	\label{eq:delta_F_4}
	\delta\mathcal{F}_4(\beta)[u]=\int_0^1\rho\left(\beta_0'+\frac{1}{\rho}\cos\beta_0\sin\beta_0\right)^3\left(u'+\frac{u}{\rho}\right)\dd\rho
\end{equation}  
is unbounded below. This can easily be seen by computing $\delta\mathcal{F}_4(\beta)[u]$ for $u_A=A\rho$, with $A$ a constant, and taking the limit as $A\to-\infty$. Thus, we have proved that \eqref{eq:beta_perturbation_0} can hold only for $\alpha=2$ and that both $\beta_\lambda$ and $\mathcal{F}_\lambda[\beta_\lambda]$ differ from $\beta_0$ and $\mathcal{F}_\lambda[\beta_\lambda]$, respectively, by terms in $\lambda^2$, as confirmed by the graphs in Figs.~\ref{fig:beta_profiles} and \ref{fig:energy_profile}.

We now consider the limit as $\lambda\to\infty$. In this case, we start by remarking that the uniform state represented by $\beta\equiv0$ is a solution of \eqref{eq:EL_bulk_clcs} for all values of $\lambda$. As in \eqref{eq:beta_perturbation_0}, we perturb it by setting 
\begin{equation}
	\label{eq:beta_perturbation_infinity}
	\beta_\lambda(\rho)=\frac{1}{\lambda^\alpha}u(\rho),
\end{equation}
where again $\alpha>0$ and $u\in\mathcal{H}^{1}([0,1])\cap \mathcal{L}^\infty([0,1])$ such that $u(0)=0$. To determine $\alpha$ and $u$, we minimize $\mathcal{F}_\lambda[\beta_\lambda]$, which for $\lambda\gg1$ is given the approximate form
\begin{equation}
	\label{eq:F_lambda_infinity}
	\mathcal{F}_\lambda[\beta_\lambda]\approx\frac12\frac{1}{\lambda^{2\alpha}}\delta^2\mathcal{F}_2(0)[u]+\frac{1}{24}\frac{1}{\lambda^{4\alpha-2}}\delta^4\mathcal{F}_4(0)[u],
\end{equation}
where $\delta^4\mathcal{F}_4$ is the fourth variation of $\mathcal{F}_4$, the first that does not vanish identically on $\beta\equiv0$.

For $\alpha>1$, $1/\lambda^{2\alpha}$ is the dominant power in \eqref{eq:F_lambda_infinity}, but the minimum of $\mathcal{F}_\lambda$ is not attained because $\delta^2\mathcal{F}_2(0)$ is unbounded below. To see this, it suffices to compute 
\begin{equation}
	\label{eq:delta_2_F_2_computation}
	\delta^2\mathcal{F}_2(0)[u]=\int_0^1\left(\rho u'^2+\frac{u^2}{\rho}\right)\dd\rho+(1-2k_{24})u^2(1)
\end{equation}
for $u_A=A\rho$, with $A$ again a positive constant, which gives
\begin{equation}
	\label{eq:delta_2_F_2_unbounded}
	\delta^2\mathcal{F}_2(0)[u_A]=2A^2(1-k_{24})\to-\infty\quad\text{for}\quad A\to\infty,
\end{equation}
by \eqref{eq:scaled_constants_inequalities}. 

For $\alpha=1$, both terms in \eqref{eq:F_lambda_infinity} scale like $1/\lambda^2$ and $\mathcal{F}_\lambda$ becomes
\begin{equation}
	\label{eq:F_lambda_infinity_become}
	\mathcal{F}_\lambda[\beta_\lambda]\approx\frac{1}{2\lambda^2}\left\{\int_0^1\left[\rho u'^2+\frac{u^2}{\rho}+\frac12\rho\left(u'+\frac{u}{\rho}\right)^4\right]\dd\rho+(1-2k_{24})u^2(1) \right\},
\end{equation}
whose equilibrium equations are
\begin{subequations}
	\begin{align}
		\frac{1}{\rho}u-u'-\rho u''&=0,\label{eq:EL_infty}\\
		\left(\rho u'+\frac{1}{\rho^2}(\rho u'+u)^3+(1-2k_{24})u\right)_{\rho=1}&=0,
	\end{align}
\end{subequations}
which are solved by $u= u_\infty$ with
\begin{equation}
	\label{eq:u_infty}
	u_\infty(\rho):=\frac12\sqrt{k_{24}-1}\rho.
\end{equation}

For $\alpha<1$, the second power in \eqref{eq:F_lambda_infinity} prevails on the first, and the minimum is trivially $u\equiv0$. Thus, we conclude that in \eqref{eq:beta_perturbation_infinity} we must set $\alpha=1$ and $u= u_\infty$, that is,
\begin{equation}
	\label{eq:beta_lambda_approximate}
	\beta_\lambda(\rho)\approx \frac{1}{2\lambda}\sqrt{k_{24}-1}\rho.
\end{equation}
Correspondingly, it follows from \eqref{eq:F_lambda_infinity_become} that 
\begin{equation}
	\label{eq:free_sup_infty}
	\mathcal{F}_\lambda[\beta_\lambda]\approx-\frac{1}{8}(k_{24}-1)^2.
\end{equation}

Finally, we study, for any given $\lambda$, the behaviour of the solutions to \eqref{eq:EL_bulk_clcs} subject to \eqref{eq:boundary_condition_0} for $\rho\ll1$. To this end, we take $\beta_\lambda(\rho)\approx A\rho^\alpha$ with $A$ and $\alpha$ positive constants and compute the limit as $\rho\to0$ in \eqref{eq:EL_bulk_clcs}; it is easily checked that compatibility demands that $\alpha=1$. Similarly, by taking $\beta_\lambda(\rho)\approx A\rho+B\rho^2$ and computing again the limit as $\rho\to0$ in \eqref{eq:EL_bulk_clcs}, we see that $B$ must vanish. We conclude that the minimum $\beta_\lambda$ of $\mathcal{F}_\lambda$ has the following asymptotic behaviour,
\begin{equation}
	\label{eq:asymptotic_beta_lambda}
	\beta_\lambda(\rho)=A\rho+\order(\rho^3),
\end{equation} 
with $A$ a constant, in agreement with the graphs shown in Fig.~\ref{fig:beta_profiles}. The constant $A$ is related to the twist $T$ on the cylinder's axis, as it follows from \eqref{eq:asymptotic_beta_lambda} and \eqref{n_bur_characteristics_T} that for $\rho=0$ the distortion characteristics are 
\begin{equation}
	\label{eq:distortion_characteristics_axis}
	S=B=q=0\quad\text{and}\quad T=\frac{2A}{R},
\end{equation}
typical of a double twist.

\section{Conclusions}\label{sec:conclusion}
The study of chromonics, which are lyotropic liquid crystals resulting from supramolecular aggregations dissolved in water, has revealed the emergence at the macroscopic scale of a twisted ground state, which owing to the achiral nature of the microscopic constituents occurs equally likely in two variants of opposite chiralities.

To accommodate these new materials within the classical elastic theory of liquid crystals, researchers hypothesized anomalously small twist constants and (comparatively) large saddle-splay constants, violating one of the Ericksen inequalities, which guarantee the classical Oseen-Frank stored energy to be bounded below.
For equilibrium problems in confined, fixed geometries, such a violation  seemed to have no consequences on the existence and stability of energy minimizers, but free-boundary problems were found paradoxical. In particular, minimizing sequences of tactoidal droplets were shown to drive the total energy to negative infinity by squeezing (double) twist within ever growing needle-shaped domains.

This paper proposes to describe the elasticity of chromonics by a adding to the Oseen-Frank classical energy density a term quartic in the measure of twist.   
In so doing, the theory introduces a phenomenological length $a$, whose microscopic origin is as problematic as the origin of spontaneous (single) twist in cholesteric liquid crystals. We may speculate that $a$ depends on temperature and it possibly arises from the flexibility of the supramolecular aggregates that constitute these materials.

Two are the major conclusions reached here. First, the proposed quartic twist theory  is proven to remove the paradoxes encountered in free-boundary problems with the classical quadratic theory.

Second, when applied to chromonics confined in a capillary tube enforcing degenerate planar anchoring for the nematic director $\n$ on the lateral boundary, the quartic twist theory predicts equilibrium configurations for $\n$ that depend on the ratio $\lambda=a/R$, where $R$ is the capillary's radius. Using published experimental data for SSY, we can estimate both $K_{24}$ and $a$. To further put the theory to the test, it might be worthwhile performing similar experiments on a series of capillaries with very different radia.

Two other theoretical challenges are suggested by this study. First, the proposed theory posits a degenerate double twist as ground state for chromonics. As recently shown in \cite{long:explicit} for the non-degenerate double twist in the ground state of cholesterics, this is a source of elastic frustration. A natural question is then how such a frustration could be eased; the answer should depend on how the domain length scale compares with the phenomenological length $a$.

A second, more demanding question is how $a$ is related to the structure of the microscopic constituents of the material. 
\nigh{A possible conjecture is that $a$ arises from fluctuations in shape of the columnar aggregates that constitute the material. A statistical mechanics model that accounts for serpentining modes generated by the gliding of each molecular disk over its neighbors might identify the length scale associated with the most likely mode; this would then be related to $a$.}

These difficult questions should be addressed by future research.  

\appendix
\nigh{
	\section{Data fitting}\label{sec:fits}
	In the experiment performed in \cite{davidson:chiral} that we try to interpret with our theory, the capillary's radius  $R=45.3\,\mu\mathrm{m}$ was determined within an error $\Delta R=0.4\,\mu\mathrm{m}$. Thus, for a given $a$, two different values of $\lambda$ would be associated with $R^\pm:=R\pm\Delta R$,
	\begin{equation}
		\label{eq:lambda_+_-}
		\lambda^\pm:=\frac{\lambda}{1\pm\Delta R/R}.
	\end{equation}
	Actually, since $\Delta R/R\approx8.8\times10^{-3}$, these values are close to one another and $\lambda^+/\lambda^-\approx1.02$. For a number of selected radii $0<r_i<R$, the polar angle $\beta$ was measured, producing data $\beta_i$, each of which can be associated with any $\rho^+_i<\rho<\rho^-_i$, where
	\begin{equation}
		\label{eq:rho_i_+_-}
		\rho_i^\pm:=\frac{\rho_i}{1\pm\Delta R/R}\quad\text{with}\quad\rho_i:=\frac{r_i}{R}.
	\end{equation}
	Hence the red bar representing each $\beta_i$ in Fig.~\ref{fig:estimate}.
	
		\begin{figure}[h]
		\centering 
		\includegraphics[width=0.6\linewidth]{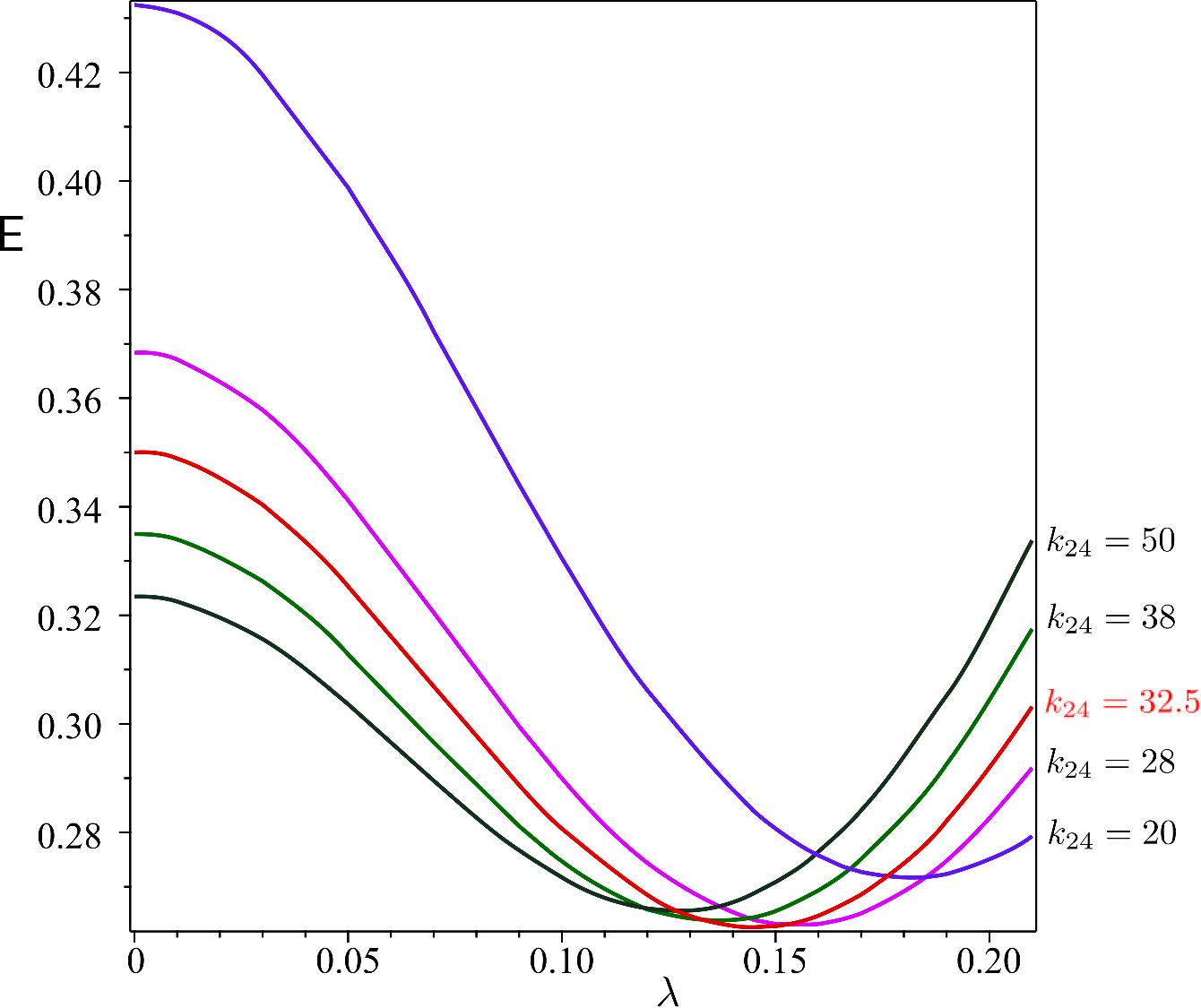}
		\caption{\nigh{Several plots of $\mathsf{E}$ against $\lambda$ for $k_3=8.7$ and selected values of $k_{24}$ illustrating the attainment of the minimum error for $k_{24}=32.5$ and $\lambda=0.142$.}}
		\label{fig:fitting}
	\end{figure}
	We took $k_3=8.7$ and used $\lambda$ and $k_{24}$ as fitting parameters and defined the (relative) error $\mathsf{E}$ to be minimized as
	\begin{equation}
		\label{eq:error_definition}
		\mathsf{E}(\lambda,k_{24}):=\frac{\sum_i\left[(\beta_{\lambda^-}(\rho_i^-)-\beta_i)^2+(\beta_{\lambda^+}(\rho_i^+)-\beta_i)^2\right]}{\int_0^1\beta_\lambda(\rho)^2\dd \rho},
	\end{equation}
	where the sum is extended to all collected data. In Fig.~\ref{fig:fitting}, $\mathsf{E}$ is plotted as a function of $\lambda$ for several values of $k_{24}$.
	The minimum of $\mathsf{E}$ is attained for $k_{24}=32.5$ and $\lambda=0.142$; its value is $\mathsf{E}_\mathrm{min}=0.263$. The corresponding function $\beta_\lambda$ is plotted against the data bars in Fig.~\ref{fig:estimate} (red curve). By setting $\lambda=0$ and seeking the minimum of $\mathsf{E}$ in \emph{both} $k_3$ and $k_{24}$, we found $k_3=6.67$ and $k_{24}=27.5$, in complete agreement with \cite{davidson:chiral}. For this choice of parameters $\mathsf{E}_\mathrm{min}=0.254$. The corresponding function $\beta_0$ is plotted against the data in Fig.~\ref{fig:estimate} (blue curve); it is barely distinguishable from $\beta_\lambda$.
}

%\bibliography{Chromonics}

%

\end{document}